\documentclass[preprint,showpacs,preprintnumbers,amsmath,amssymb]{revtex4}
\usepackage{threeparttable}
\usepackage{graphicx}
\usepackage{pifont}
\usepackage{textcomp}
\usepackage{rotating}
\usepackage{dcolumn}
\usepackage{multirow}
\begin{document}

\title{ Accurate and Efficient Solution of the Electronic Schr{\"o}dinger Equation with the Coulomb Singularity by the Distributed Approximating Functional Method}

\author{Zhigang Sun$^a$}
\affiliation{State Key Laboratory of Molecular
Reaction Dynamics and Center for Theoretical Computational Chemistry,
Dalian Institute of Chemical Physics, Chinese
Academy of Sciences, Dalian, P.R. China 116023; Center for Advanced Chemical Physics and 2011 Frontier Centre for Quantum Science and Technology, University of Science and Technology of China, 96 Jinzhai Road, Hefei 230026, China.}

\begin{abstract}
We proposed a distributed approximating functional method for efficiently describing the electronic dynamics in atoms and molecules in the presence of the Coulomb singularities, using the kernel of a grid representation derived by using the solutions of the Coulomb differential equation based upon the Schwartz's interpolation formula, and a grid representation using the Lobatto/Radau shape functions. The elements of the resulted Hamiltonian matrix are confined in a narrow diagonal band, which is similar to that using the (higher order) finite difference methods. However, the spectral convergence properties of the original grid representations are retained in the proposed distributed approximating functional method for solving the Schr{\"o}dinger equation involving the Coulomb singularity. Thus the method is effective for solving the electronic Schr{\"o}dinger equation using iterative methods where the action of the Hamiltonian matrix on the wave function need to evaluate many times. The method is investigated by examining its convergence behaviours for calculating the electronic states of the H atom, H$_2^+$ molecule, the H atom in a parallel magnetic and electric fields, as the radial basis functions.

\vfill

\noindent  $^a$ To whom corresponding should be addressed: Email:zsun@dicp.ac.cn

\end{abstract}

\maketitle
\section{INTRODUCTION}
Along with the technological development of ultra-short laser pulses, solving the time-dependent Schr{\"o}dinger equation (TDSE) describing electronic dynamics attracts more and more interest. The solution of TDSE presents accurate prediction of the laser-atom/molecule interaction. However, limited by the theoretical methods and computational resources, few systems can be simulated accurately by the TDSE in full dimensionality.  In addition, with $(t, t')$ method,  solution of a TDSE can be transformed as solving a time-independent Schr{\"o}dinger equation (TISE).\cite{yao1994,peskin1994} From this aspect, the numerical methods for solving TDSE is of similar interest in solving a TISE and vice cersa.

In the solution of a TISE for finding bound or resonance states, the iterative Lanczos and its variants are optimal choices when grid representation is adopted since it leads to favourable computational scalings.\cite{saad2003,guo2007} Iterative methods involve repeated actions of a Hamiltonian matrix, $\hat{H}$, on a wave function or wave packet represented as a vector $\Psi$ of function values at the grid points. The efficiency of the evaluation of the repeated $\hat{H}\Psi$ products determine the computational speed. Similarly, in the solution of a TDSE, the initial wavefunction is advanced by an evolution operator which, if the Hamiltonian $\hat{H}$ is time independent,  is an exponential function $\hat{U}(t; \hat{H}) = \exp(i\hat{H}t)$ (in atomic units), or if the Hamiltonian $\hat{H}$ is time dependent, can be approximated by a product of a series of exponential functions $\hat{U}(t; \hat{H}) = \exp(i\hat{H}\Delta t)\exp(i\hat{H}\Delta t)\dots$ by ignoring the time-ordering.\cite{lef1991} The former can be approximated by, for example, a polynomial expansion, and the later short time propagator can be evaluated by the split operator or Lanczos method etc.\cite{lef1991} In any case,  the basic operation is also reduced to the evaluation of the action of the Hamiltonian operator onto the wave packet, $\hat{H}\Psi$. In a long time propagation this operation has to be repeated many times, and thus, similar to that for solving a TISE, its evaluation is the computational bottleneck. In a solution of the Schr{\"o}dinger equation with time-independent Hamiltonian using real Chebyshev wavepacket method, the difference between "time-dependent" or "time-independent" method completely disappears, where the time-evolution becomes simple Chebyshev polynomial expansion involving repeated evaluation of $\hat{H}\Psi$ but without "time" parameter.\cite{chen1996} 

In the field of chemical dynamics, there have been well development in numerical grid methods for solving the Schr{\"o}dinger equation. One of the major techniques is the general discrete variable representation (DVR) method, which was purposed in 1985 by Light et al. \cite{light1985,miller1992,light2000} The main advantages of the DVR method is that the resulted Hamiltonian matrix is very sparse thus it is particularly suitable for solving the nuclear Shr{\"{o}}dinger equation by iterative methods. \cite{kosloff1988,lef1991}  Currently many DVRs have appeared in the literatures.\cite{light2000,muckerman1990,miller1992,stare2003,dhzhangh2plusoh}
The other popular methods, such as the Lagrange mesh method (LMM) and quadrature discretization method (QDM) for solving the Schr{\"o}dinger equation, share much spirit with the DVR method.\cite{Lo2006,Baye1999}  In a calculation adopted the DVR method, considered only local operators this matrix would be diagonal, but the Hamiltonian includes the kinetic energy operator which is nonlocal in the coordinate representation. If the number of grid points in each coordinate $\alpha$ is $n_{\alpha}$, the Hamiltonian matrix will contain of the order of $N\times n$ nonzero elements, where $N$ is the total number of grid points $N=\Pi_{\alpha}n_{\alpha}$ and $n=\sum_{\alpha}n_{\alpha}$. Apparently, sparser Hamiltonian matrix will lead to faster evaluation of $\hat{H}\Psi$ and less numerical effort.

Regarding with  the Coulomb singularity, which arises in describing the electronic dynamics of atoms and molecules, however, extensive studies indicate
that most of the current DVR methods do not work well using the Gauss quadrature rule.\cite{shimshovitz2012} Only a few of them are good for treating the Coulomb singularity.
The DVR of the generalised Laguerre  polynomials only is able to describe
a subset of the eigenstate of a Coulomb potential  at a time and they are not suitable for describing the ionisation continua.\cite{weaver1992,boyd2003}
The Lobatto-DVR can accurately represent the Coulomb singularity in spherical coordinates, with unnecessarily dense grid points at both ends of the grid,\cite{schneider2004,Yu2015} where the Lobatto shape functions on Gauss-Lobatto quadrature are taken as the basis.
\cite{manolopoulos1988,schneider2004,guan2011}
The Coulomb DVR works well with a single Coulomb singularity in spherical coordinates,
where its corresponding basis is the Coulomb wave functions. \cite{dunseath2002,peng2006} 

The Coulomb potential is a long range potential, which extends far away from the nuclei. To describe the dynamics of electrons induced by laser pulses, such as near threshold IR and high frequency XUV ionization, usually grid in a long range (thus many grid points) is required. The DVR and its decedent methods are global methods, which usually are utilised with classical basis function and have spectral convergence, thus leading to full matrix of large $n_{\alpha}$ for one particular degree of freedom $\alpha$.  On the other hand, the Hamiltonian matrix, constructed by the local methods, such as the finite difference (FD) and finite element (FE) methods,\cite{rescigno2000,dqyu2013} is sparser as banded matrix. Therefore, with the same number grid points, the action of the Hamiltonian matrix on the wavefunction with the local method can be evaluated faster.  Unfortunately, the convergence speed of low order FD or FE method is slower than the global method, and usually more grid points are required. $B$-spline method is quite often adopted for solving a TISE/TDSE involving Coulomb potential, however, the resulted Hamiltonian matrix is not so sparse as that using the FD or DVR method.\cite{bachau2001,vanroose2006}

The distributed approximating functional (DAF) method presents a good solution to this problem.\cite{Hoffman1991,Hoffman1992} It can be combined with a kind of kernels,\cite{Wei1997,Wei1999} and leads to banded Hamiltonian matrix but with spectral convergence. For finding vibrational states of a diatomic Morse potential, it have been proven that the DAF with the Sinc DVR kernel is inferior to the sum acceleration FD or spectral FD method, due to their better weighted cardinal functions.\cite{Mazziotti1999,boyd2006,wei2007}  However, the DAF is easy to accomplish with different kernels, even with non-evenly distributed grid points, and usually retains the spectral convergence of its kernel with a suitable width. 

In this work, we propose a new class of DAF, which is constructed by combining with the Lobatto DVR (LDVR) or Radau DVR (RDVR) and Coulomb DVR (CDVR), which have been proven being able to treat well with the Coulomb singularity, with rapid decreasing weight functions $\mathrm{w}(r,\sigma)$. We will show that these DAFs work excellently for figuring out the electronic states in the presence of the Coulomb singularity, similar to their kernels, but result in well banded Hamiltonian matrix, thus are very appealing in an effective numerical solution of a TISE or TDSE. 

The content of the remaining paper is arranged as following: Section II presents the theoretical details for the proposed DAFs; In Section III, the new DAFs were illustrated by finding the electronic bound states of hydrogen atom in spherical coordinates, the bound states of H$_2^+$ molecule in spherical protate coordinates, and the resonance states of hydrogen atom in spherical coordinates in parallel magnetic and electric fields using the complex scaling method. The DAFs are compared with the traditional higher order FD methods for calculating bound states of hydrogen atom. In the calculations, the DAFs are taken as the basis functions of the radial degree of freedom. Section IV concludes the present work.

\section{Theoretical Methods}
Following Peng and Starace and Dunseath et al.,\cite{peng2006,dunseath2002}  with simple zeros at the real points $r_j$ of the analytic reference function $v(r)$, which is the solution of 
the Coulomb differential equation
\begin{equation}
\left[ \frac{d^2}{dr^2}-\frac{\lambda(\lambda+1)}{r^2}+\frac{2Z}{r}+2E\right]v(r)=0,\label{coulomb}
\end{equation}
the Coulomb DVR functions can be defined as 
\begin{equation}
f_i(r)=\frac{1}{\sqrt{\omega_i}}C_i(r)=\frac{1}{\sqrt{\omega_i}}\frac{1}{v'(r_i)}\frac{v(r)}{r-r_i}
\end{equation}
where $\sqrt{\omega_i}$ is defined as $\frac{\sqrt{\pi}}{v'(r_i)}$. These functions satisfy
\begin{equation}
f_{i}(r_{j})=\frac{\delta_{ij}}{\sqrt{\omega_{i}}},
\end{equation}
and are orthogonal
\begin{equation}
\int^{a}_{b}f_{i}(r)f_{j}(r)dr\simeq\sum_{m=1}^{N}\omega_{m}f_{i}(r_{m})f_{j}(r_{m})=\delta_{ij}.
\end{equation}
Using these functions as basis, the coordinate operator is diagonal, in the spirit of the DVR,
\begin{equation}
\int^{a}_{b}f_{i}(r)rf_{j}(r)dr\simeq\sum_{m=1}^{N}\omega_{m}f_{i}(r_{m})r_{m}f_{j}(r_{m})=\delta_{ij}r_{i}.
\end{equation}

In the CDVR, the Hamiltonian matrix of the Schr{\"o}dinger equation can be derived using the following relations
\begin{equation}
P_{ij}=\int^{\infty}_0 f^*_i(r)\frac{d}{dr}f_i(r)dr=(1-\delta_{ij})\frac{1}{r_i-r_j}
\end{equation}
and
\begin{equation}
T_{ij}=-\int^{\infty}_0 f^*_i(r)\frac{d^2}{dr^2}f_i(r)dr=-\delta_{ij}\frac{c_i}{3a_i}+(1-\delta_{ij})\frac{2}{(r_i-r_j)^2}
\end{equation}
with $a_i$ and $c_i$ are defined as $v'(r_i)$ and  $a_i(-2E-\frac{2Z}{r_i})$, respectively. In practice, usually parameters $E$ and $Z$ in Eq.\ref{coulomb} are adjustable but $\lambda$ is taken as zero.
This CDVR  can accurately represent both bound and continuum wave functions of electrons along the radial degree of freedom in a Coulomb potential using spherical coordinates. However, the Hamiltonian matrix 
$P_{ij}$ and $T_{ij}$ are dense and full, which leads to large computational effort when the radial degree of freedom has to be long range in practice.  

Similar to the Lagrange distributed approximating functionals (LDAF) proposed by Wei and his co-workers for solving the Schr{\"o}dinger equation for molecular motion and Fokker-Planck equation,\cite{Wei1997,Wei1999} the DAF with the CDVR functions as kernel may be defined as 
\begin{equation}
\phi_i(r,\sigma)=f_i(r)\mathrm{w}(r,\sigma)
\end{equation}
with $\mathrm{w}(r,\sigma)$ defined as
$$\mathrm{w}(r,\sigma)=\exp[-(r-r_i)^2/2\sigma^2]$$
which by decaying to zero prevents derivatives of $\phi_i(r,\sigma)$, named as CDAF, at $r_j$ from coupling with points separated long from the grid point $r_i$. The parameter $\sigma$ controls the decaying rate.
With the CDAF, the Hamiltonian matrix can be written as
\begin{equation}
\bar{P}_{ij}=P_{ij}\mathrm{w}_j(r_i,\sigma)+\delta_{ij}\mathrm{w}'_j(r_i,\sigma)
\end{equation}
and
\begin{equation}
\bar{T}_{ij}=T_{ij}\mathrm{w}_j(r_i,\sigma)+2P_{ij}\mathrm{w}'_j(r_i,\sigma)+\delta_{ij}\mathrm{w}''_j(r_i,\sigma)
\end{equation}

Similarly, the DVR with the Lobatto/Radau shape functions, can be taken as the kernel of the DAF also, which applies the Gauss-Lobatto/Radau quadrature and was put forward by Manolopoulos and Wyatt.\cite{manolopoulos1988}
The LDVR basis functions, which are actually the Lagrangian interpolating polynomials, are given by
\begin{equation}\label{eq:lobattofunctions}
u_{i}(r)=\frac{1}{\sqrt{w_{i}}}\prod_{j=0}^{N+1}\frac{r-r_{j}}{r_{i}-r_{j}},j\neq i, i=0,\ldots,N+1.
\end{equation}
Clearly, these functions satisfy:
\begin{equation}
u_{i}(r_{j})=\frac{\delta_{ij}}{\sqrt{w_{i}}},
\end{equation}
Furthermore, they are orthogonal under the Gauss-Lobatto/Radau quadrature rule,
\begin{equation}
\int^{a}_{b}u_{i}(r)u_{j}(r)dr\simeq\sum_{m=0}^{N+1}w_{m}u_{i}(r_{m})u_{j}(r_{m})=\delta_{ij}.
\end{equation}
Similarly, using these functions as basis, the coordinate operator is diagonal,
\begin{equation}
\int^{a}_{b}u_{i}(r)xu_{j}(r)dr\simeq\sum_{m=0}^{N+1}w_{m}u_{i}(r_{m})r_{m}u_{j}(r_{m})=\delta_{ij}r_{i}.
\end{equation}

In the LDVR, the Hamiltonian matrix $T_{ij}$  can be written as
\begin{equation}
T_{ij}=\sum\limits_{k}\frac{df_{i}(r_{k})}{dr}\frac{df_{j}(r_{k})}{dr}w_{k},
\end{equation}
and
\begin{equation}
P_{ik}=\frac{df_{i}(r_{k})}{dr}=
\left\{
\begin{array}{lll}
-0.5/w_{k},\qquad\qquad\qquad\qquad\qquad i=1,k=1\\
0.5/w_{k},\qquad\qquad\qquad\qquad\qquad\;\;\; i=N,k=N\\
(1-\delta_{ik})\frac{1}{r_{i}-r_{k}}\prod\limits_{l\neq m,l\neq k}\frac{r_{k}-r_{l}}{r_{i}-r_{l}^{i}}, \qquad\; \text{else}
\end{array}
\right.
\end{equation}

Same as the CDAF, the DAF with LDVR/RDVR functions as the kernel (LODAF/RDVR) then may be defined as 
\begin{equation}
\Phi_i(r,\sigma)=u_i(r)\mathrm{w}(r,\sigma).
\end{equation}
In the LODAF/RDVR, the Hamiltonian matrix can be similarly derived as that in the CDAF.

Both in the CDAF and LODAF/RDAF, the cardinal functions are weighted before differentiation, thus the formulas for the accelerated derivatives may contain more terms than in the spectral FD method.\cite{Mazziotti1999} However, the CDAF and LODAF can
be implemented straight forward, and of spectral convergence same as its precedent DVR method as we will show below for solving the electronic Schr{\"o}dinger equation involving the Coulomb singularity. Its simplicity, robustness efficiency are intriguing.

\section{Results and Discussion}
\subsection{Bound states of hydrogen atom}

The non-relativistic Hamiltonian of the hydrogen atom is written as
\begin{equation}
H=-\frac{1}{2}\frac{d^2}{dr^2}+\frac{L(L+1)}{r^2}-\frac{1}{r}
\end{equation}
Here $L=0$ was applied for a tough numerical test for dealing with the Coulomb singularity. The grid extends to 3200 atomic unit (a.u.), in order to support accurately the 30th state of hydrogen atom. In the LODAF calculation, the number of grid points adopted is 240, but in the CDAF calculation, $E$=0.001 a.u. and $Z$=8.0 are adopted, which leads to 152 grid points. These parameters are just capable of giving eigenvalues of the supported bound states with accuracy of machine accuracy (double precision). Apparently, as comparing with the CDVR method, the unnecessary dense grid points at both of the grid ends of the LDVR reduce its numerical efficiency. This problem can be alleviated by using the recently proposed mapped DVR scheme.\cite{dqyu2013,Yu2015} 

Besides the above calculations, 1200 grid points for LODAF and $E$=0.1 a.u. and $Z$=12.0 for CDAF, which lead to total 503 grid points, were also adopted to check the numerical properties of $\sigma$, the width of weighting function, with the grid points of a denser distribution.

The numerical convergences of several typical electronic states with respect to $\sigma$, the width of weighting function, are presented in Fig.\ref{hydrogen}, where the error is defined as
\begin{equation}
{\rm error}=\frac{|E_i-E^0_i|}{E^0_i}.
\end{equation}
$E_i$ is the eigenvalule calculated by using the DAF method, however, $E_i^0$ is the "standard" eigenvalule calculated using the corresponding DVR method with the exactly same parameters. The eigenenergies of the Hamiltonian matrix in all of the calculations in the present work are finding either by the direct diagonalization method or by the Lanczos method and its variants, depending on the size of the Hamiltonian matrix.

With increasing $\sigma$, the numerical results converge rapidly for both the LODAF and CDAF, particularly for the low lying states, as seen in the panels of Fig.\ref{hydrogen}. Higher lying states require $\sigma$ of larger values to converge. This is because that the last hump of higher lying state extends to large radial coordinate where the grid points are sparser or the wave function oscillates more rapidly, as shown in the left panel of Fig.\ref{wfm}. Thus the grid points couple with each other in a longer range and only $\sigma$ of larger values guarantee an accurate description. With increasing grid points, the width $\sigma$ for obtaining converged results correspondingly decreases, as shown in Fig.\ref{hydrogen}(C) and (D), which reflects the fixed bandwidth of the states. 

Comparing with the whole grid range, quite small value of $\sigma$ is required for obtaining converged results, which thus leads to Hamiltonian matrix of very limited band width, as schematically shown in the right panel of Fig.\ref{wfm}. For the CDAF with $\sigma$=20a.u. in the case with 503 grid points, the band width is about 60 elements at the beginning of the grid, and the band width is only about 35 elements at the end of the grid, if the matrix elements of value less than $1.0\times10^{-10}$ are eliminated. The band of the matrix is well limited. Thus, this feature is of particular advantage for  a TDSE calculation or finding eigestates with the Lanczos type method, where the matrix of the Hamiltonian need act on the wave function many times thus saving much computational effort as comparing with the original DVR method.

\subsection{Bound states of H$_2^+$ molecule}
It is convenient to describe the electronic dynamics of a diatomic molecule in spherical prolate coordinates.  The angular degree of freedom is represented by the Legendre DVR method. For the radial degree of freedom, it has been proven that the DVR using the Radau shape functions is a good choice.\cite{tao2009} Similar to the LDVR, the  DAF can be realised with the Radau DVR kernel. The Hamiltonian for describing the dynamics of the electron of H$^+_2$  can be written as
\begin{eqnarray}
H=&-&\frac{1}{2a^2(\xi^2-\eta^2)}\left(\frac{\partial}{\partial\xi}(\xi^2-1)\frac{\partial}{\partial\xi}+\frac{\partial}{\partial\eta}(\eta^2-1)\frac{\partial}{\partial\eta}\right. \nonumber\\
 &-&\left.\frac{m^2}{\xi^2-1}-\frac{m^2}{1-\eta^2}+2a(Z_1+Z_2)\xi+2a(Z_1-Z_2)\eta   \right)\label{h2p}
\end{eqnarray}
For the meaning of the variables in the equations, one may refer Ref.\citenum{tao2009}. By Eq.\ref{h2p}, the DAF with the kernel of RDVR can been implemented straight forward.  

In the calculations, the grid range for $\xi$ is set as (1.0, 120.0]. $Z_1=-Z_2$ is set as 1.0 and $m$=0.  The number of grid points for $\xi$ is 120 and for $\eta$ is 20, respectively.  The eigenstates of H$_2^+$ below 25th state converge well with these parameters.  The convergence behaviours of typical states, whose eigenvalues obtained using the corresponding DVR method are listed in Tab.\ref{tab1} and taken as the standard reference values, as a function of $\sigma$ are plotted in the left panel of Fig.\ref{hh}.  It is observed that even for the 25th state, with $\sigma$=4.0, we can obtain its eigenvalue with accuracy of error less than 10$^{-13}$. With such small value of $\sigma$, the Hamiltonian matrix again has very limited band width for the radial degree of freedom, which would save much computational effort in a computation where $H\psi$ is required to evaluate many times, as comparing with the Radau DVR method.  

\subsection{Resonance states of hydrogen atom in parallel magnetic and electric fields}

The behaviours of a hydrogen atom in external electric and magnetic fields is a basic quantum mechanical problem, which has been studied over a century due to its fundamental significance. \cite{schweizer2002} 
Various variational techniques and analytic solution of hydrogen atom in a homogeneous magnetic field have been reported. However, those calculations are not so easy to perform and the corresponding wavefunction are often heavy.\cite{kra1996,stubbins2004,dimova2005,zhao2006,main1992,cart2007} Simple but accurate and efficient numerical methods for a solution of hydrogen atom in magnetic and electric field are still of interest.\cite{baye2008,wunner2014,mele1993,yxzhang2008,rao1995,fass1996,guan2006,guan2004} At the same time, due to its fundamental significance, many numerical methods have been applied for solving the hydrogen atom in magnetic and electric fields, thus solution of this problem provides a good prototype for illustrating the properties of a new numerical method. \cite{baye2008,wunner2014,yxzhang2008,rao1995,sahoo2001,jent2001,zhao2007}

The non-relativistic Hamiltonian of hydrogen atom in parallel magnetic and electric fields can be written as \cite{rao1995}
\begin{eqnarray}
H=&&-\frac{d^2}{2dr^2}+\frac{1}{2r^2}\left[ \frac{d}{d\cos\theta}(1-\cos^2\theta)\frac{d}{d\cos\theta}+\frac{m^2}{1-\cos^2\theta} \right]\nonumber\\
&&+\frac{1}{2}m\gamma-\frac{1}{r}+\frac{1}{8}\gamma^2(1-\cos^2\theta)r^2+fr\cos\theta
\end{eqnarray}
where the parameter $\gamma$ expresses the magnetic field $\mathbf{B}$ in atomic units of $B_0 = \hbar/a^2_0e \approx 2.35 \times 10^5$T and electric field strength $f=F/F_0$ with $F_0\approx5.14\times10^9$V/cm. Here the proton mass is assumed to be infinite.  The eigenstates of  Hamiltonian of the hydrogen atom in a magnetic field in $z$ direction are solved using the proposed LODAF or CDAF for radial degree of freedom and the Legendre DVR adopting symmetry for the angular degree of freedom. 

\subsubsection{hydrogen atom in a strong magnetic field}

With $r \in [0, 150]$ a.u., 120  grid points in the LDVR are adopted for $r$ but 75 grid points for $\theta$ of the Legendre DVR are adopted using the symmetry, the eigenvalues of the first five states with $\gamma$=1 are listed in Tab.\ref{tab1} and taken as the standard reference values. The convergence of these five eigenvalues as a function of $\sigma$ using the LODAF, are plotted in the right panel of Fig.\ref{hh}.  We see again that the results converge to accuracy of error less than 10$^{-9}$ with $\sigma$ as small as 3.0 a.u. The Hamiltonian matrix for radial coordinate has very limited band width again using the LODAF in this case.

\subsubsection{hydrogen atom in  parallel magnetic and electric fields}

The quantum states of hydrogen atom in a combined electric and magnetic fields appear as resonant states due to the coupling between continuum and bound states. Thus one need proper boundary conditions in a calculation with limited grid extension. One of the most effective methods is the complex rotation method, which mathematically is rigorous, unlike the complex absorbing potential method usually adopted in a time-dependent wavepacket calculations.\cite{ho1983,nissen2010,kalita2011} After complex scaling $r\to re^{i\phi}$, a non-Hermitian Hamiltonian is obtained as
\begin{eqnarray}
H=&&-e^{-2\phi}\frac{d^2}{2dr^2}+e^{-2i\phi}\frac{1}{2r^2}\left[ \frac{d}{d\cos\theta}(1-\cos^2\theta)\frac{d}{d\cos\theta}+\frac{m^2}{1-\cos^2\theta} \right]\nonumber\\
&&+\frac{1}{2}m\gamma-\frac{e^{-i\phi}}{r}+\frac{1}{8}\gamma^2(1-\cos^2\theta)r^2e^{2i\phi}+fr\cos\theta e^{i\phi}
\end{eqnarray}

Both the CDVR and LDVR and their DAFs are taken as the basis function for radial coordinate in the following calculations. Since there is no detailed investigation for this problem using the CDVR and LDVR for the radial coordinate in the literatures, the eigenenergy of the ground state with different basis sets (grid point numbers)  of the LDVR with $\gamma$=1$\times 10^{-1}$ and $f$=2$\times10^{-1}$ are listed in Tab.\ref{tab2}, along with the results reported in Ref.\citenum{yxzhang2008} using the $B$-spline basis.  The extent of the radial coordinate is [0.0, 15.0]. It can be seen that the numerical convergence is very rapid with increasing number of grid points. Only with basis set of size as $N_r\times N_{\theta}$ = 20$\times$10, the resonance energy $E_{\rm res}$ can be obtained with 11 significant digits. 

To illustrate the efficiency of the CDAF and LODAF, higher excited resonance states $n=10$ are investigated with $\gamma$=2$\times 10^{-4}$ and $f$=1.4$\times10^{-5}$ , which requires the grid range to extend to $r_{\rm max}\approx$ 600.0 in order to acquire accurate results.  Two different sizes of the basis sets were applied in the calculations. One is small but large just enough to give converged results. The other one is much larger for the basis sets for the radial coordinate, in order to reproduce the conditions in a realistic calculation including ionization continua.  The results are given in Tab.{\ref{tab3}, using both the LDVR and CDVR, and comparing with the results reported in Ref.\citenum{rao1995}. It is seen that in the grid range of such large extent, the resonance energies converge well with very limited grid points. The results in Tab.\ref{tab3} are taken as the standard values to examine the convergence behaviours with respect to the width $\sigma$ of weighting of the CDAF and LODAF. 

The convergence behaviours of resonance energies and lifetimes for ($n_1$, $n_2$)=(0, 9), (1, 8), and (2, 7), and also (0, 0) state as a function of $\sigma$  for the CDAF with basis sets as 174$\times$13 and 245$\times$13, and for the LODAF with basis sets as 90$\times$13 and 240$\times$15, are shown in Fig.\ref{fig4}. For radial coordinate extending to 600.0 a.u., the CDAF with width small as 10.0 a.u. can give accurate results using small basis set as shown in Fig.4(A), and using large basis set, the width $\sigma$ for obtaining converged results is even smaller as 6.0 a.u. as shown in Fig.4(C). Similar convergence behaviours are seen in the calculations using the LODAF, as shown in panels (B) and (D). These are very encouraging results, since the Hamiltonian matrix for the radial coordinate becomes much sparser comparing with that using the original DVR method. The resonance energy $E_{\rm res}$ and the lifetime $\Gamma$ of a particular state exhibit same behaviours in the calculations, and for brevity, only convergence behaviours of one of them are presented.

\subsection{Comparison between the Higher-order FD and DAFs for Calculating Bound states of H atom}

As we have noticed in the right panel of Fig.\ref{wfm}, the structure of the Hamiltonian matrix in the DAF is very similar to that in a calculation using the higher order or spectral FD method \cite{Mazziotti1999,thachuk1992,gray2001,guantes1999,mazz2002,farnum2004}. We also notice that the weighting width $\sigma$ decreases with denser grid points in a converged calculation, as we have shown in Fig.\ref{hydrogen} and Fig.\ref{fig4}. The required values of $\sigma$ in converged calculations almost are inversely proportional to the number of the grid points. Smaller grid spacing requires the calculation using the grid points extending in a narrower space. This feature is similar to the higher oder FD method also. It seems that the FD method and the DAF method perhaps have close relationship with each other. At the same time, the FD method is popular for treating the Coulomb singularities problem, especially in a TDSE calculation\cite{peng2006,dundas2000,curdy2001,yuan2011} and atomic structure calculations.\cite{fischer1976} Thus, it wold be interesting to compare the numerical convergence of the DAF and various FD methods for treating with the Coulomb potential.

The FD(2n) methods use function values at a number of points to obtain an estimate of the second derivative of the function at a particular point, which $2n$ indicate the order of the FD method. The general formula for the estimation of the second derivative of a function $p(x)$ at $x_0$, using a grid of $N=2n+1$ points given by
\begin{equation}
x_k=x_0+k\Delta x, \quad k=0,\pm1,\dots,\pm n
\end{equation}
is\cite{miller1992}
\begin{equation}
p^{\prime\prime}(x_0)=-\frac{1}{\Delta x^2}[2p_0\sum_{l=1}^n\frac{1}{l^2}-\sum_{k=1}^n(p_k+f_{-k})\frac{1}{k^2}\prod_{l=1}^{n}\null^{\prime}(\frac{l^2}{l^2-k^2})]
\end{equation}
The grid is laid out so that grid point(s) lies around boundary $r=0$ requiring the values of the wave function at $r_k < 0$ when such higher order centred FD methods are applied. Therefore, for these grid points, the one-side FD method is adopted in the following calculations with the same order of the centred finite difference for interior grid points.\cite{fornberg1996} This will lead to asymmetric Hamiltonian matrix and results in "ghost" states, but with high accuracy for calculating the true eigenstates.\cite{zhao2007}  Sometimes one may also modified certain coefficients of the higher order finite difference method, in order to obtain better eigenenergy of the ground state.\cite{peng2006,dundas2000} This method is not applied here since we are investigating the numerical convergence behaviours of the FD method, instead of a realistic calculation. 

The energies of the ground state of hydrogen atom given by the FD method of different orders, as a function of the numbers of grid points, are presented in Tab.\ref{tab4} with radial coordinate in the range of  [0, 18.0]. Higher order and more grid points give results of better accuracy, as expected. Using the FD(18) method, 60 grid points are capable of giving the energy with accuracy of 10 significant digits. It is interesting to see that the FD method with higher order than 18 does not improve the accuracy of the ground state anymore using 60 grid points. The energies of the ground state of hydrogen atom calculated in the same grid range using the LDVR method but with 1/10 of the numbers of the grid points, are given in Tab.\ref{tab4} as the last column for comparison.  To obtain accurate energy of the ground state, only 20 grid points of LDVR are required, which is much less than those required in the FD method.

With grid range of [0, 3200.0], the numerical convergence of 1st, 5th and 15th state calculated by FD(2), FD(4), $\dots$, FD(40) as a function of the number grid points are presented in Fig.\ref{fig5}(A), (B) and (C), respectively. The results calculated using the Sinc DVR method are presented also for comparison.\cite{miller1992} It is seen that higher order FD method converges faster, and higher excited state is easier to converge which is less sensitive to the Coulomb singularity. And the Sinc DVR is worse than all of the FD methods for calculating electronic states of hydrogen atom, due to the presence of the singularity. It is also interesting to see that the highest accuracy for the high order finite difference methods, such as FD(40) or FD(36) etc., is quite low, only about 10$^{-6}$ a.u., especially for the 15th excited state as shown in Fig.\ref{fig5}(C). This is perhaps due to the mixed usage of the one-side finite difference for the grid points around $r=0$ and centred finite difference for interior grid points. 

The numerical convergence behaviours of the LDVR in the same grid range [0.0, 3200.0]a.u. for the 1st, 5th, 10th, 15th, 20th and 25th state are presented in Fig.\ref{fig5}(D). Comparing with the results in the other three panels, it is clear that the LDVR method converges much faster than the FD methods. At the same time, "ghost" states do not arise in the LDVR calculations with any number of grid points, unlike in the calculations using the FD methods. 

On the other hand, since in the FD method the grid points distribute evenly, in contrast with that the grid points concentrate around the ends of the grid range in the LDVR.  Therefore, such comparison between the numerical efficiencies of the FD methods and LDVR method is not fair. It is very possible that the FD methods, in a combination with certain variable mapped functions, would become much more efficient, similar to that for the Sinc DVR and Fourier method.\cite{fattal1996,modine1997,kawata1999,boyd2003,Lin2015} In that case the spectral FD methods may be very useful also for treating with the Coulomb potential.\cite{boyd1991,boyd1994,Mazziotti1999} This worth more investigation in the future work. 

A practical issue in a calculation is that, if the efficiency of grid points of the DVR methods has been deteriorated by a combination with the DAF formalism, or if "ghost" states arise like that in a calculation using the FD method. As shown in panel (A) of Fig.\ref{fig6}, the convergence behaviours of typical states, 1st, 10th and 20th, as a function of the number of grid points using the LODAF with $\sigma$=100.0 a.u., are very similar to those using the LDVR method. With $\sigma$=60.0, the convergence of 20th state is a little slower, but reach machine accuracy with the 200 grid points. This is expected from Fig.\ref{hydrogen}(B). The numerical convergence of the eigen energies of lowest 50 states with different $\sigma$ using the CDAF are presented in Fig.\ref{fig6}, which exhibit similar behaviors. Since the the grid range, which is the same as that  used for Fig.1 and only supports the lowest about 30 states, the difference, which is defined as $\log_{10}|E^0_i-E_i|$, between the energies obtained by the CDAF ($E_i$) and CDVR ($E_i^0$) of these 30 states decrease rapidly with increasing $\sigma$. However, the difference of the states, which have energies above the asymptotic energy,  does not change much with increasing $\sigma$. This is due to the different boundary conditions in the CDAF and CDVR. In a practical calculation where the ionization continua have to be considered, we alway need impose suitable absorbing boundary conditions and such convergence problem for the continua would disappear. Anyway, "ghost" states does not arise in a DAF calculation. Similar numerical behaviours happen with the LODAF and LDVR. Thus the original spectral convergence of the LDVR/CDVR method is indeed retained in the LODAF/CDAF method.  

The Lobatto/Radau DVR and the CDVR method deal with the continua and bound states with the same weight.\cite{peng2006,schneider2004} Therefore, the conclusion of above analysis applies to the case where ionization continua need to be considered. Especially, the wavelength of continua decreases with increasing ionization energy, which requires smaller $\sigma$ in a converged calculation and lead to sparser Hamiltonian matrix. Thus inclusion of ionization continua may not introduce more difficulty than the numerical examples above. The merit of the introduced methods are of particularly interest for describing the electronic dynamics of an atom or a molecule induced by ultra-short laser pulses with long wavelength, i.e., describing the near threshold ionization processes induced by IR pulses, where grid points in a long range (thus many grid points) are required. In any case, the DAF can reduce the number of the Hamiltonian matrix elements drastically thus saves computational effort. In the future work this issue will be investigated with realistic numerical experiments.

In the above calculations, all numerical examples only involves single electron. In many-electron problems, there are two-electron (electron-electron repulsion) operators which are local. The present DAF scheme does nothing to reduce the number of 2-body integrals, which are by far more numerous than one-body terms. By choosing suitable coordinate system and variable mapping schemes, similar DAF methods with some kernel may be developed which can approximate the electron-electron repulsion with good accuracy.  This is out from the scope of the present work.

\section{Conclusion}

The distributed approximating functionals with the Lobatto/Radau DVR and CDVR kernels are introduced for treating with the Coulomb singularities in atoms and molecules, which results in the Hamiltonian matrix to be of similar shape to that using the higher-order finite difference method thus very sparse. These DAF approaches is extremely simple and entails low CPU cost due to its slow scaling with problem size for solving the electronic Schr{\"o}dinger equation. At the same time, it is of spectral convergence and free of "ghost" states, essentially same as its precedent DVR method. This is in contrast with the finite element DVR method and (higher order) finite difference method. The method should be of particular interest in a calculation for solving the Schr{\"o}dinger equation using iterative methods, where the action of the Hamiltonian matrix on the wave function need to evaluate many times. With this method, it may be solvable to accurately investigate the electronic dynamics in near threshold ionization and the effect to the eletronic re-collision processes of the high Rydberg states using the current computational resources.

{\bf Acknowledgements}
This work was supported by the National Basic Research Program of China
(973 program, No. 2013CB922200 \& 2012YQ1200470403), the National Natural Science Foundation of China
(Grant No. 21222308, 21103187, and 21133006), the Chinese Academy of Sciences and the Key Research Program of the 
Chinese Academy of Sciences.

\bibliographystyle{jcp}


\begin{thebibliography}{10}

\bibitem{yao1994}
{\sc G.~H. Yao} and {\sc R.~E. Wyatt},
\newblock {\em J. Chem. Phys.} {\bf 101}, 1904 (1994).

\bibitem{peskin1994}
{\sc U.~Peskin}, {\sc R.~Kosloff}, and {\sc N.~Moiseye},
\newblock {\em J. Chem. Phys.} {\bf 100}, 8849 (1994).

\bibitem{saad2003}
{\sc Y.~Saad},
\newblock {\em Iterative Methods for Sparse Linear Systems: Second Edition},
\newblock SIAM, Philadelphia, 2003.

\bibitem{guo2007}
{\sc H.~Guo},
\newblock {\em Reviews in Computational Chemistry, Volume 25}, chapter 7,
  Recursive Solutions to Large Eigenproblems in Molecular Spectroscopy and
  Reaction Dynamics,
\newblock Wiley-VCH, 2007.

\bibitem{lef1991}
{\sc C.~Leforestier}, {\sc R.~H. Bisseling}, {\sc C.~Cerjan}, {\sc M.~D. Feit},
  {\sc R.~Friesner}, {\sc A.~Guldberg}, {\sc A.~Hammerich}, {\sc G.~Jolicard},
  {\sc W.~Karrlein}, {\sc H.-D. Meyer}, {\sc N.~Lipkin}, {\sc O.~Roncero}, and
  {\sc R.~Kosloff},
\newblock {\em J. Comput. Phys.} {\bf 94}, 59 (1991).

\bibitem{chen1996}
{\sc R.~Chen} and {\sc H.~Guo},
\newblock {\em J. Chem. Phys.} {\bf 105}, 3569 (1996).

\bibitem{light1985}
{\sc J.~C. Light}, {\sc I.~P. Hamilton}, and {\sc J.~V. Lill},
\newblock {\em J. Chem. Phys.} {\bf 82}, 1400 (1985).

\bibitem{miller1992}
{\sc T.~D. Colbert} and {\sc W.~H. Miller},
\newblock {\em J. Chem. Phys.} {\bf 96}, 1982 (1992).

\bibitem{light2000}
{\sc J.~C. Light} and {\sc T.~{Carrington Jr.}},
\newblock {\em Adv. Chem. Phys.} {\bf 114}, 263 (2000).

\bibitem{kosloff1988}
{\sc R.~Kosloff},
\newblock {\em J. Phys. Chem.} {\bf 92}, 2087 (1988).

\bibitem{muckerman1990}
{\sc J.~T. Muckerman},
\newblock {\em Chem. Phys. Lett.} {\bf 173}, 200 (1990).

\bibitem{stare2003}
{\sc J.~Stare} and {\sc G.~G. Balint-Kurti},
\newblock {\em J. Phys. Chem. A} {\bf 107}, 7204 (2003).

\bibitem{dhzhangh2plusoh}
{\sc D.~H. Zhang} and {\sc J.~C. Light},
\newblock {\em J. Chem. Phys.} {\bf 106}, 551 (1997).

\bibitem{Lo2006}
{\sc J.~Lo} and {\sc B.~D. Shizgal},
\newblock {\em J. Chem. Phys.} {\bf 125}, 194108 (2006).

\bibitem{Baye1999}
{\sc D.~Baye} and {\sc M.~Vincke},
\newblock {\em Phys. Rev. E} {\bf 59}, 7195 (1999).

\bibitem{shimshovitz2012}
{\sc A.~Shimshovitz} and {\sc D.~J. Tannor},
\newblock {\em J. Chem. Phys.} {\bf 137}, 101103 (2012).

\bibitem{schneider2004}
{\sc B.~I. Schneider} and {\sc N.~Nygaard},
\newblock {\em Phys. Rev. E} {\bf 70}, 056706 (2004).

\bibitem{weaver1992}
{\sc R.~V. Weaver}, {\sc J.~T. Muckerman}, and {\sc T.~Uzer},
\newblock {\em Time-dependent Quantum Molecular Dynamics}, chapter An Analytic
  Discrete Variable Representation for the Coulomb Problem,
\newblock Plenum Press, New York, 1992.

\bibitem{boyd2003}
{\sc J.~P. Boyd}, {\sc C.~Rangan}, and {\sc P.~H. Bucksbaum},
\newblock {\em J. Comput. Phys.} {\bf 188}, 56 (2003).

\bibitem{Yu2015}
{\sc D.~Q. Yu}, {\sc S.~L. Cong}, and {\sc Z.~G. Sun},
\newblock {\em Chem. Phys.} {\bf 458}, 41 (2015).

\bibitem{manolopoulos1988}
{\sc D.~E. Manolopoulos} and {\sc R.~E. Wyatt},
\newblock {\em Chem. Phys. Lett.} {\bf 152}, 23 (1988).

\bibitem{guan2011}
{\sc X.~Guan}, {\sc K.~Bartschat}, and {\sc B.~I. Schneider},
\newblock {\em Phys. Rev. A} {\bf 83}, 043403 (2011).

\bibitem{dunseath2002}
{\sc K.~M. Dunseath}, {\sc J.-M. Launay}, {\sc M.~Terao-Dunseath}, and {\sc
  L.~Mouret},
\newblock {\em J. Phys. B: At. Mol. Opt.} {\bf 35}, 3539 (2002).

\bibitem{peng2006}
{\sc L.~Y. Peng} and {\sc A.~F. Starace},
\newblock {\em J. Chem. Phys.} {\bf 125}, 154311 (2006).

\bibitem{rescigno2000}
{\sc T.~N. Rescigno} and {\sc C.~W. McCurdy},
\newblock {\em Phys. Rev. A} {\bf 62}, 032706 (2000).

\bibitem{dqyu2013}
{\sc D.~Q. Yu}, {\sc S.~L. Cong}, {\sc D.~H. Zhang}, and {\sc Z.~G. Sun},
\newblock {\em Chin. J. Chem. Phys.} {\bf 112}, 755 (2013).

\bibitem{bachau2001}
{\sc H.~Bachau}, {\sc E.~Cormier}, {\sc P.~Decleva}, {\sc J.~E. Hansen}, and
  {\sc F.~Mart{\'i}n},
\newblock {\em Rep. Prog. Phys.} {\bf 64}, 1815 (2001).

\bibitem{vanroose2006}
{\sc W.~Vanroose}, {\sc D.~A. Horner}, {\sc F.~Mart{\'i}n}, {\sc T.~N.
  Rescigno}, and {\sc C.~W. McCurdy},
\newblock {\em Phys. Rev. A} {\bf 74}, 052702 (2006).

\bibitem{Hoffman1991}
{\sc D.~K. Hoffman}, {\sc N.~Nayar}, {\sc O.~A. Sharafeddin}, and {\sc D.~J.
  Kouri},
\newblock {\em J. Phys. Chem.} {\bf 95}, 8299 (1991).

\bibitem{Hoffman1992}
{\sc D.~K. Hoffman} and {\sc D.~J. Kouri},
\newblock {\em J. Phys. Chem.} {\bf 96}, 1179 (1992).

\bibitem{Wei1997}
{\sc G.~W. Wei}, {\sc D.~S. Zhang}, {\sc D.~G. Kouri}, and {\sc D.~K. Hoffman},
\newblock {\em Phys. Rev. Lett.} {\bf 79}, 775 (1997).

\bibitem{Wei1999}
{\sc G.~W. Wei},
\newblock {\em J. Chem. Phys.} {\bf 110}, 8930 (1999).

\bibitem{Mazziotti1999}
{\sc D.~A. Mazziotti},
\newblock {\em Chem. Phys. Lett.} {\bf 299}, 473 (1999).

\bibitem{boyd2006}
{\sc J.~P. Boyd},
\newblock {\em J. Comp. Phys.} {\bf 214}, 538 (2007).

\bibitem{wei2007}
{\sc G.~W. Wei} and {\sc S.~Zhao},
\newblock {\em J. Comp. Phys.} {\bf 226}, 2389 (2007).

\bibitem{tao2009}
{\sc L.~Tao}, {\sc C.~W. McCurdy}, and {\sc T.~N. Rescigno},
\newblock {\em Phys. Rev. A} {\bf 79}, 012719 (2009).

\bibitem{schweizer2002}
{\sc P.~Schmelcher} and {\sc W.~Schweizer},
\newblock {\em Atoms and Molecules in Strong External Fields},
\newblock Klumer, New York, 2002.

\bibitem{kra1996}
{\sc Y.~P. Kravchenko}, {\sc M.~A. Liberman}, and {\sc B.~Johansson},
\newblock {\em Phys. Rev. Lett.} {\bf 77}, 619 (1996).

\bibitem{stubbins2004}
{\sc C.~Stubbins}, {\sc K.~Das}, and {\sc Y.~Shiferaw},
\newblock {\em J. Phys. B: At. Mol. Opt. Phys.} {\bf 37}, 2201 (2004).

\bibitem{dimova2005}
{\sc M.~G. Dimova}, {\sc M.~S. Kaschiev}, and {\sc S.~Vinitsky},
\newblock {\em J. Phys. B: At. Mol. Opt. Phys.} {\bf 38}, 2337 (2005).

\bibitem{zhao2006}
{\sc L.~B. Zhao} and {\sc P.~C. Stancil},
\newblock {\em Phys. Rev. A} {\bf 74}, 055401 (2006).

\bibitem{main1992}
{\sc J.~Main} and {\sc G.~Wunner},
\newblock {\em Phys. Rev. Lett.} {\bf 69}, 586 (1992).

\bibitem{cart2007}
{\sc H.~Cartarius}, {\sc J.~Main}, and {\sc G.~Wunner},
\newblock {\em Phys. Rev. Lett.} {\bf 99}, 173003 (2007).

\bibitem{baye2008}
{\sc D.~Baye}, {\sc M.~Vincke}, and {\sc M.~Hesse},
\newblock {\em J. Phys. B: At. Mol. Opt. Phys} {\bf 41}, 055005 (2008).

\bibitem{wunner2014}
{\sc C.~Schimeczek} and {\sc G.~Wunner},
\newblock {\em Comp. Phys. Comm.} {\bf 185}, 614 (2014).

\bibitem{mele1993}
{\sc V.~S. Melezhik},
\newblock {\em Phys. Rev. A} {\bf 48}, 4528 (1993).

\bibitem{yxzhang2008}
{\sc Y.~X. Zhang}, {\sc H.~Y. Meng}, and {\sc T.~Y. Shi},
\newblock {\em Chin. Phys. B} {\bf 17}, 140 (2008).

\bibitem{rao1995}
{\sc J.~G. Rao} and {\sc B.~W. Li},
\newblock {\em Phys. Rev. A} {\bf 51}, 4526 (1995).

\bibitem{fass1996}
{\sc P.~Fassbinder} and {\sc W.~Schzeizer},
\newblock {\em Phys. Rev. A} {\bf 53}, 213 (1996).

\bibitem{guan2006}
{\sc X.~X. Guan},
\newblock {\em Phys. Rev. A} {\bf 74}, 023413 (2006).

\bibitem{guan2004}
{\sc X.~X. Guan} and {\sc Y.~X. Zhang},
\newblock {\em Phys. Rev. A} {\bf 71}, 033409 (2004).

\bibitem{sahoo2001}
{\sc S.~Sahoo} and {\sc Y.~K. Ho},
\newblock {\em Phys. Rev. A} {\bf 65}, 015403 (2001).

\bibitem{jent2001}
{\sc U.~D. Jentschura},
\newblock {\em Phys. Rev. A} {\bf 64}, 013403 (2001).

\bibitem{zhao2007}
{\sc L.~B. Zhao} and {\sc P.~C. Stancil},
\newblock {\em J. Phys. B: At. Mol. Opt. Phys.} {\bf 40}, 4347 (2007).

\bibitem{ho1983}
{\sc Y.~K. Ho},
\newblock {\em Phys. Rep.} {\bf 99}, 1 (1983).

\bibitem{nissen2010}
{\sc A.~Nissen}, {\sc H.~O. Karlsson}, and {\sc G.~Kreiss},
\newblock {\em J. Chem. Phys.} {\bf 133}, 054306 (2010).

\bibitem{kalita2011}
{\sc D.~J. Kalita} and {\sc A.~K. Gupta},
\newblock {\em J. Chem. Phys.} {\bf 134}, 094301 (2011).

\bibitem{thachuk1992}
{\sc M.~Thachuk} and {\sc G.~C. Schztz},
\newblock {\em J. Chem. Phys.} {\bf 97}, 7297 (1992).

\bibitem{gray2001}
{\sc S.~K. Gray} and {\sc E.~M. Goldfield},
\newblock {\em J. Chem. Phys.} {\bf 115}, 8331 (2001).

\bibitem{guantes1999}
{\sc R.~Guantes} and {\sc S.~C. Farantos},
\newblock {\em J. Chem. Phys.} {\bf 111}, 10827 (1999).

\bibitem{mazz2002}
{\sc D.~A. Mazziotti},
\newblock {\em J. Chem. Phys.} {\bf 117}, 2455 (2002).

\bibitem{farnum2004}
{\sc J.~D. Farnum} and {\sc D.~A. Mazziotti},
\newblock {\em J. Chem. Phys.} {\bf 120}, 5962 (2004).

\bibitem{dundas2000}
{\sc D.~Dundas}, {\sc J.~F. {McCann}}, {\sc J.~S. Parker}, and {\sc K.~T.
  Taylor},
\newblock {\em J. Phys. B: At. Mol. Opt.} {\bf 33}, 3261 (2000).

\bibitem{curdy2001}
{\sc M.~Baertschy}, {\sc T.~N. Rescigno}, {\sc W.~A. Isaacs}, {\sc X.~Li}, and
  {\sc C.~W. {McCurdy}},
\newblock {\em Phys. Rev. A} {\bf 63}, 022712 (2001).

\bibitem{yuan2011}
{\sc K.~J. Yuan}, {\sc H.~Z. Lu}, and {\sc A.~D. Bandrauk},
\newblock {\em Phys. Rev. A} {\bf 83}, 043418 (2011).

\bibitem{fischer1976}
{\sc C.~F. Fischer},
\newblock {\em The Hartree-Fock Method for Atoms: A Numerical Approach},
\newblock John Wiley \& Sons, New York, 1976.

\bibitem{fornberg1996}
{\sc B.~Fornberg},
\newblock {\em SIAM Rev.} {\bf 40}, 685 (1996).

\bibitem{fattal1996}
{\sc E.~Fattal}, {\sc R.~Baer}, and {\sc R.~Kosloff},
\newblock {\em Phys. Rev. E} {\bf 53}, 1217 (1996).

\bibitem{modine1997}
{\sc N.~A. Modine}, {\sc G.~Zumbach}, and {\sc E.~Kaxiras},
\newblock {\em Phys. Rev. B} {\bf 55}, 10289 (1997).

\bibitem{kawata1999}
{\sc I.~Kawata} and {\sc H.~Kono},
\newblock {\em J. Chem. Phys.} {\bf 111}, 9498 (1999).

\bibitem{Lin2015}
{\sc X.~S. Lin} and {\sc Z.~G. Sun},
\newblock {\em Chem. Phys. Lett.} {\bf 621}, 35 (2015).

\bibitem{boyd1991}
{\sc J.~P. Boyd},
\newblock {\em Appl. Num. Math.} {\bf 7}, 287 (1991).

\bibitem{boyd1994}
{\sc J.~P. Boyd},
\newblock {\em Comput. Methods Appl. Mech. Engrg.} {\bf 116}, 1 (1994).

\bibitem{krav1996}
{\sc Y.~P. Kravchenko}, {\sc M.~A. Liberman}, and {\sc B.~Johansson},
\newblock {\em Phys. Rev. Lett.} {\bf 77}, 619 (1996).

\bibitem{telnov2007}
{\sc D.~A. Telnov} and {\sc S.~I. Chu},
\newblock {\em Phys. Rev. A} {\bf 76}, 043412 (2007).

\bibitem{kamta2005}
{\sc G.~L. Kamta} and {\sc A.~D. Bandrauk},
\newblock {\em Phys. Rev. A} {\bf 71}, 053407 (2005).

\end{thebibliography}

\newpage

\begin{table}
\caption{The standard values for evaluating the error of the DAF with the Radau DVR kernel (H$^+_2$) and the Lobatto DVR kernel (H in magnetic field) for $R=2.0$ a.u. and $m$=0, which were obtained by the corresponding DVR method. The digits in bold font indicates the converged numbers, as comparing with those reported in Ref.[\citenum{krav1996}] and [\citenum{telnov2007},\citenum{kamta2005}] }
\begin{tabular}{|c|c|c|c|c|}
\hline
\multicolumn{2}{|c|}{Energies of H$_2^+$ electronic states} &
\multicolumn{2}{c|}{Energies of H electronic states in a magnetic field}\\
\cline{1-4}
 $\quad$ 1st  $\quad$ &  $\mathbf{-1.102634214494}$ & $\quad$ 1$s_0$ $\quad$ &             $\mathbf{0.8311688967}514$ \\
\hline
5th & $\mathbf{-0.2357776288}$255 & 2$s_0$ &   $\mathbf{0.26000661594}$62 \\
\hline
10th & $\mathbf{-0.10544230117}$24 & 2$p_0$ & $\mathbf{0.1604689826}$776 \\
\hline
15th & -0.06973813856103 & 3$p_0$ & $\mathbf{0.090224511}$4576 \\
\hline
20th & -0.05567119873107 & 3$d'_0$ & $\mathbf{0.066233066}$6764 \\
\hline
25th & -0.04095255968841 &   & \\
\hline
\end{tabular}\label{tab1}
\end{table}

\begin{table}
\caption{The convergence of the $n_1=0$, $n_2$=0, $m=$0 resonance of the hydrogen atom for $\gamma$=0.1 a.u. and $f$=0.2 a.u. with the LDVR and Legendre DVR using symmetry. The values in the last column  was those reported in Ref.[\citenum{yxzhang2008}]  in 2008.}
\begin{tabular}{|c|c|l|l|l|l}
\hline
$N_r \times N_{\theta}$ & & $\alpha$=0.4 &  $\alpha$=0.5 &   $\alpha$=0.6   \\
\hline
15$\times$ 5 & $E_{\rm res}$ & -0.56804693  & -0.56804700 & -0.56804642 \\
 & $\Gamma /2$ & -5.9481578$\times 10^{-2}$ &  -5.9481653$\times 10^{-2}$ &  -5.9481857$\times 10^{-2}$  \\
\hline
20$\times$ 5 & $E_{\rm res}$& -0.56804693  & -0.56804693 & -0.56804693 \\
 & $\Gamma /2$ &  -5.94815597$\times 10^{-2}$ &  -5.94815601$\times 10^{-2}$ &-5.94815997$\times 10^{-2}$  \\
\hline
20$\times$ 10 & $E_{\rm res}$&  -0.56804590606  & -0.56804590606 & -0.56804590604 \\
 & $\Gamma /2$ &  -5.948143623$\times 10^{-2}$ &  -5.948143622$\times 10^{-2}$ &  -5.948143629 $\times 10^{-2}$  \\
\hline
Zhang et al. & $E_{\rm res}$&  -0.56804590607  & -0.56804590605 & -0.56804590604  \\
40 $\times$ 30 & $\Gamma /2$ &  -5.948143623$\times 10^{-2}$ &  -5.948143623$\times 10^{-2}$ &  -5.948143621 $\times 10^{-2}$  \\
\hline
\end{tabular}\label{tab2}
\end{table}

\begin{table}
\caption{$E_{\rm res}$ and $\Gamma$/2 for $n=10$ excited states of the hydrogen atom  for $m=0$ with $\gamma$=2$\times 10^{-4}$ a.u. and $f$=1.4$\times 10^{-5}$  a.u. with LDVR/CDVR and Legendre DVR using symmetry. The values in the last column  was those reported in Ref.[\citenum{rao1995}]  in 1995.}
\begin{tabular}{|c|c|l|l|l|l}
\hline
$N_r \times N_{\theta}$ & & $n_1$=0, $n_2$=9 & $n_1$=1, $n_2$=8 &   $n_1$=2, $n_2$=7   \\
\hline
90$\times$ 13 & $E_r$ & -0.7195855762$\times 10^{-3}$  & -6.735493456$\times 10^{-3}$ & -6.277299160$\times 10^{-3}$  \\
 LDVR & $\Gamma /2$ & 3.78461109$\times 10^{-5}$ &  1.9994530647$\times 10^{-5}$ &  8.2868377$\times 10^{-5}$  \\
\hline
240$\times$ 15 & $E_r$ & -0.7195855762$\times 10^{-3}$  & -6.735493455$\times 10^{-3}$ & -6.277299160$\times 10^{-3}$  \\
 LDVR & $\Gamma /2$ & 3.78461095$\times 10^{-5}$ &  1.9994528910$\times 10^{-5}$ &  8.2868375$\times 10^{-5}$  \\\hline
 
174$\times$ 13 & $E_r$ & -0.7195855762$\times 10^{-3}$  & -6.735493456$\times 10^{-3}$ & -6.277299160$\times 10^{-3}$  \\
 CDVR & $\Gamma /2$ & 3.78461081$\times 10^{-5}$ &  1.9994530786$\times 10^{-5}$ &  8.2868377$\times 10^{-5}$  \\\hline
245$\times$ 13 & $E_r$ & -0.7195855761$\times 10^{-3}$  & -6.735493456$\times 10^{-3}$ & -6.277299160$\times 10^{-3}$  \\
 CDVR & $\Gamma /2$ & 3.78461104$\times 10^{-5}$ &  1.9994529880$\times 10^{-5}$ &  8.2868380$\times 10^{-5}$  \\
\hline
Rao and Li & $E_r$&   -0.719585576$\times 10^{-3}$  & -6.73549346$\times 10^{-3}$ & -6.27729916$\times 10^{-3}$   \\
$l_{\rm max}$=24, $N_r$=58 & $\Gamma /2$ &  3.7846110 $\times 10^{-5}$ &  1.9994531$\times 10^{-5}$ &   8.286835 $\times 10^{-5}$  \\
\hline
\end{tabular}\label{tab3}
\end{table}

\begin{table}
\caption{The energy of ground state of hydrogen atom using the (higher order) finite difference method and LDVR with different numbers of grid points. The extent of radial coordinate is [0, 18.0]. Note the numbers of grid points used in the LDVR are 1/10 of those used in the finite difference method.}
\begin{tabular}{|c|c|l|l|l|l|l|l|}
\hline
$N_r $ & 2th & 4th & 6th &   8th & 12th &18th  & LDVR ($N_r$/10)\\
\hline
30 & 0.4637462852 & 0.4945503202 & 0.4984932382 & 0.4996528672 & 0.4999830887 & 0.4999998477 & 0.0987654321  \\\hline
60 & $0.4895653776 $ & 0.4998569108  & 0.4999779464 & 0.4999984235 & 0.4999999913 & 0.5000000000 &  0.3357312490 \\\hline
100 & $0.4960916265 $ & 0.5000034246  & 0.4999992975 & 0.4999999824 & 0.5000000000 & 0.5000000000 & 0.4952685493 \\\hline
140 & $0.4979793114 $ &0.5000034890  & 0.4999999305 & 0.4999999992 & 0.5000000000 & 0.5000000000 & 0.4999964169 \\\hline
180 & $0.4987698495 $ & 0.5000017677  & 0.4999999875 & 0.4999999999 & 0.5000000000 & 0.5000000000 & 0.4999999998 \\\hline
220 & $0.4991735177 $ & 0.5000009193  & 00.4999999968 & 0.5000000000 & 0.5000000000 & 0.5000000000 & 0.5000000000 \\\hline
\end{tabular}\label{tab4}
\end{table}

\newpage

\begin{figure}
\includegraphics[width=1.0\textwidth]{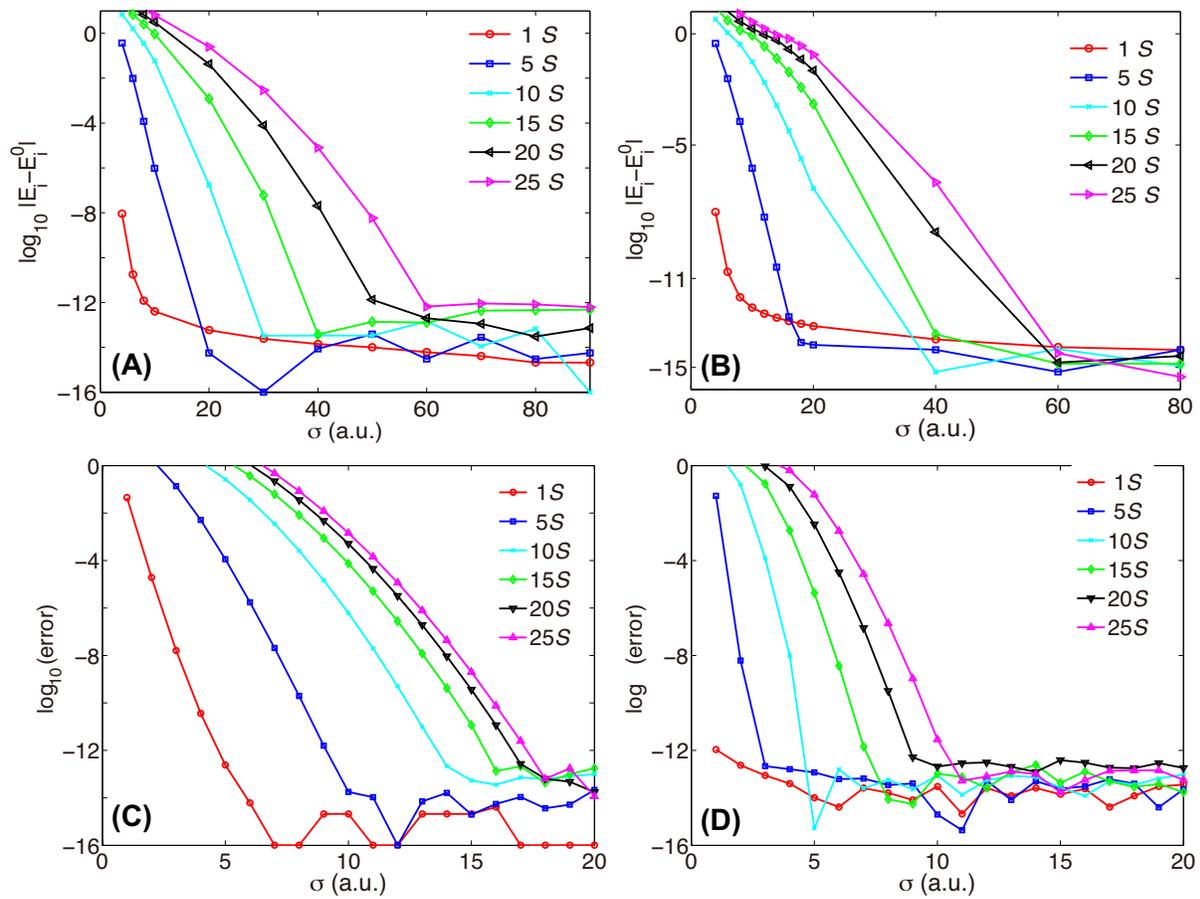}
\caption{\label{hydrogen} The convergence behaviors of energies of typical electronic states of H atom as a function of $\sigma$ for the DAF with kernel of the CDVR (A and C) and the LDVR (B and D) with small (A and B) and large basis (C and D) sets.} 
\end{figure}

\begin{figure}
\includegraphics[width=1.0\textwidth]{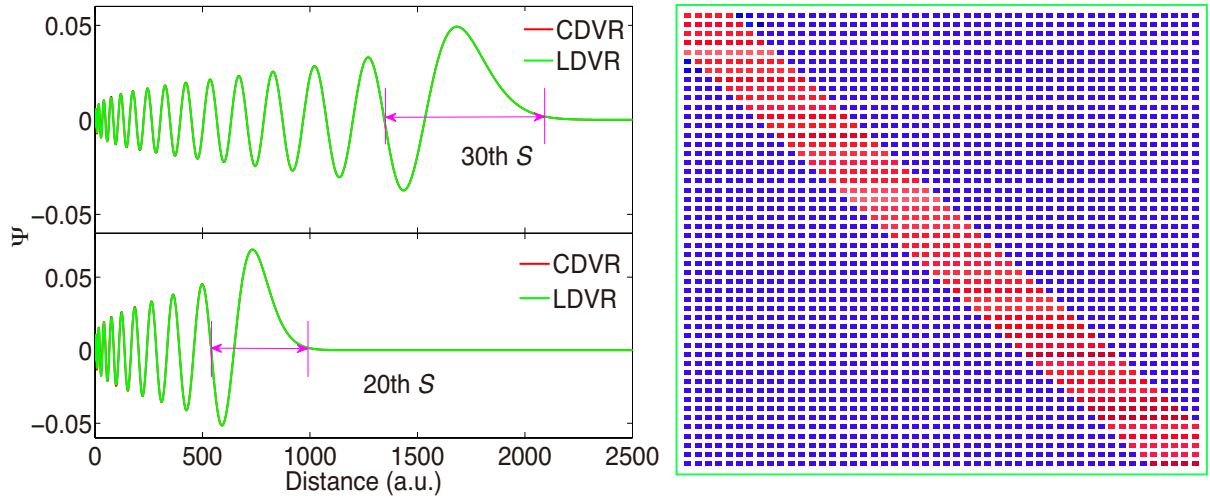}
\caption{\label{wfm} Left:  The wave functions of the 20th and 30th state of hydrogen atom using the LDVR and CDVR. Right: the matrix in the DVR method (red and blue elements) and the corresponding DAF method (red elements). The matrix using the DAF is much sparser.} 
\end{figure}

\begin{figure}
\includegraphics[width=1.0\textwidth]{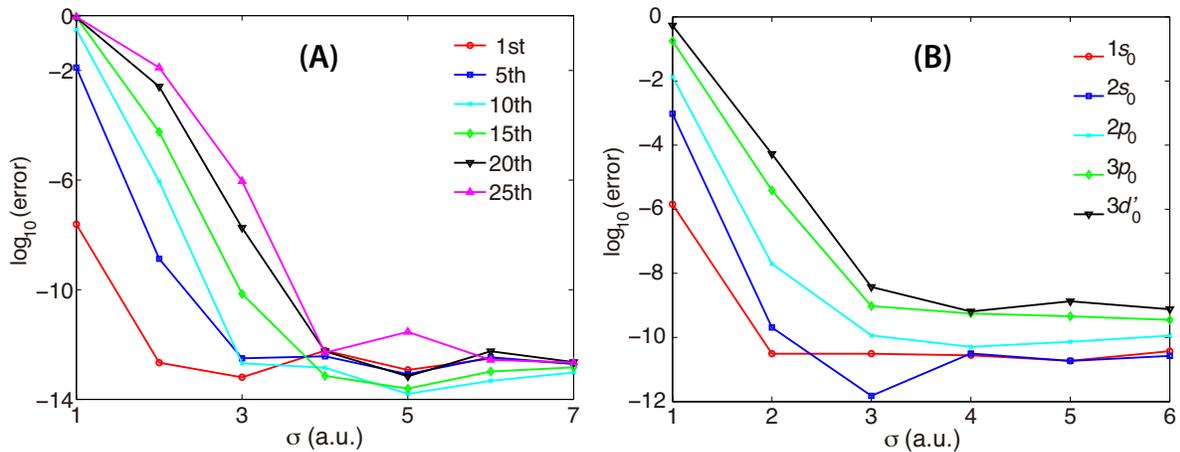}
\caption{\label{hh} Left: The convergence behaviors of energies of typical electronic states of H$_2^+$ as a function of $\sigma$ for the DAF with kernel of the RDVR. Right: The convergence behaviors of energies of lowest five electronic states of hydrogen atom in a strong magnetic field with strength of $\gamma$ = 1 a.u.} 
\end{figure}

\begin{figure}
\includegraphics[width=1.0\textwidth]{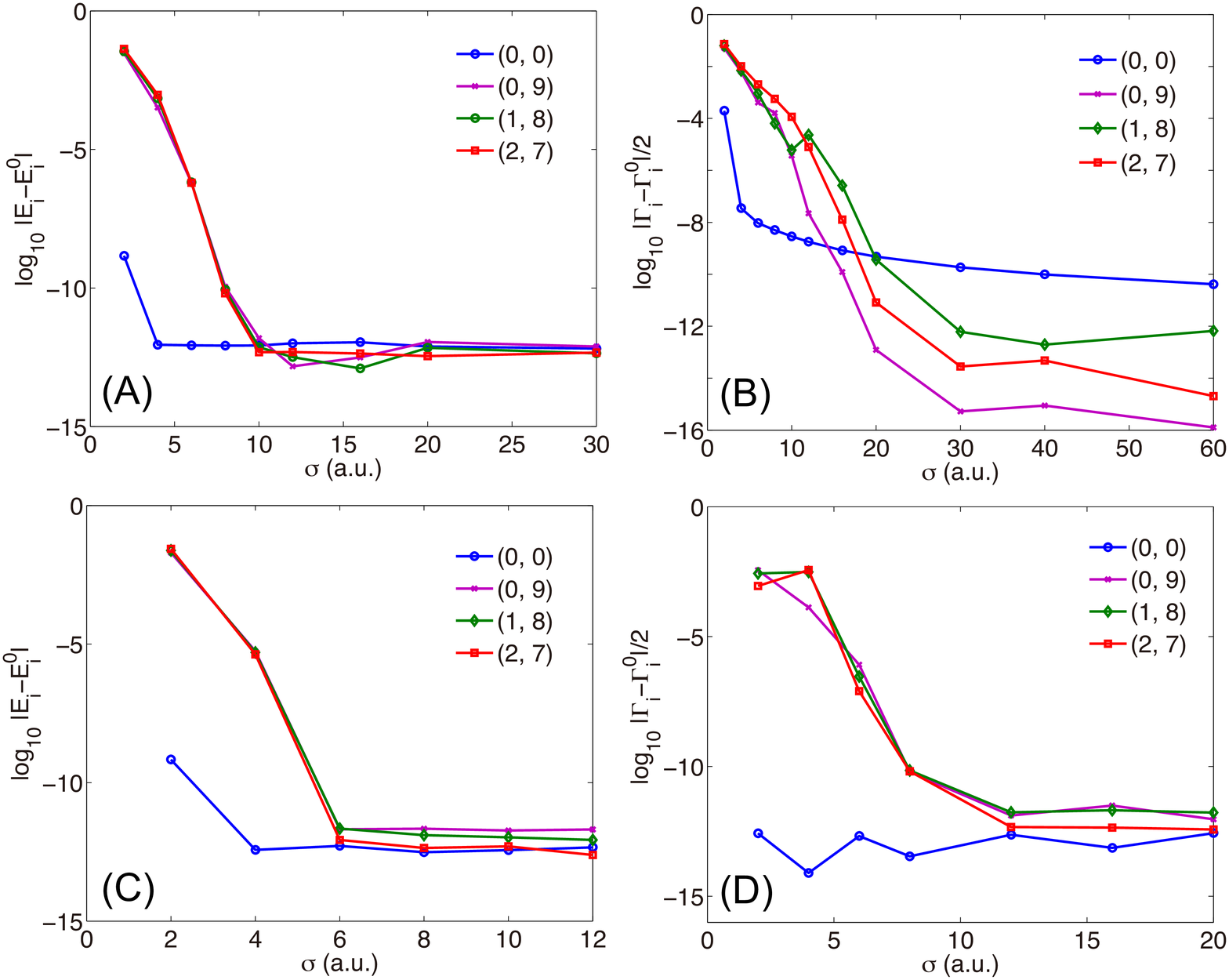}
\caption{\label{fig4} The convergence behaviours of energies and lifetimes of $n=10$ and $n=0$ resonance states of H atom in parallel magnetic and electric fields for $m=0$ with $\gamma$=2$\times 10^{-4}$ a.u. and $f$=1.4$\times 10^{-5}$  a.u., with different $\sigma$ for the DAF with kernel of the CDVR and LDVR with different basis set sizes: (A) Convergence behaviours of resonance energies $E_{\rm res}$ with basis sets as 90$\times$ 13 using the LDAF; (B): Convergence behaviours of lifetimes $\Gamma/2$ with basis sets as 240$\times$ 15 using the LDAF; (C) Convergence behaviours of resonance energies $E_{\rm res}$ with basis sets as 174 $\times$ 13 using the CDAF; (D): Convergence behaviours of lifetimes $\Gamma/2$ with basis sets as 245$\times$ 13 using the CDAF;} 
\end{figure}

\begin{figure}
\includegraphics[width=1.0\textwidth]{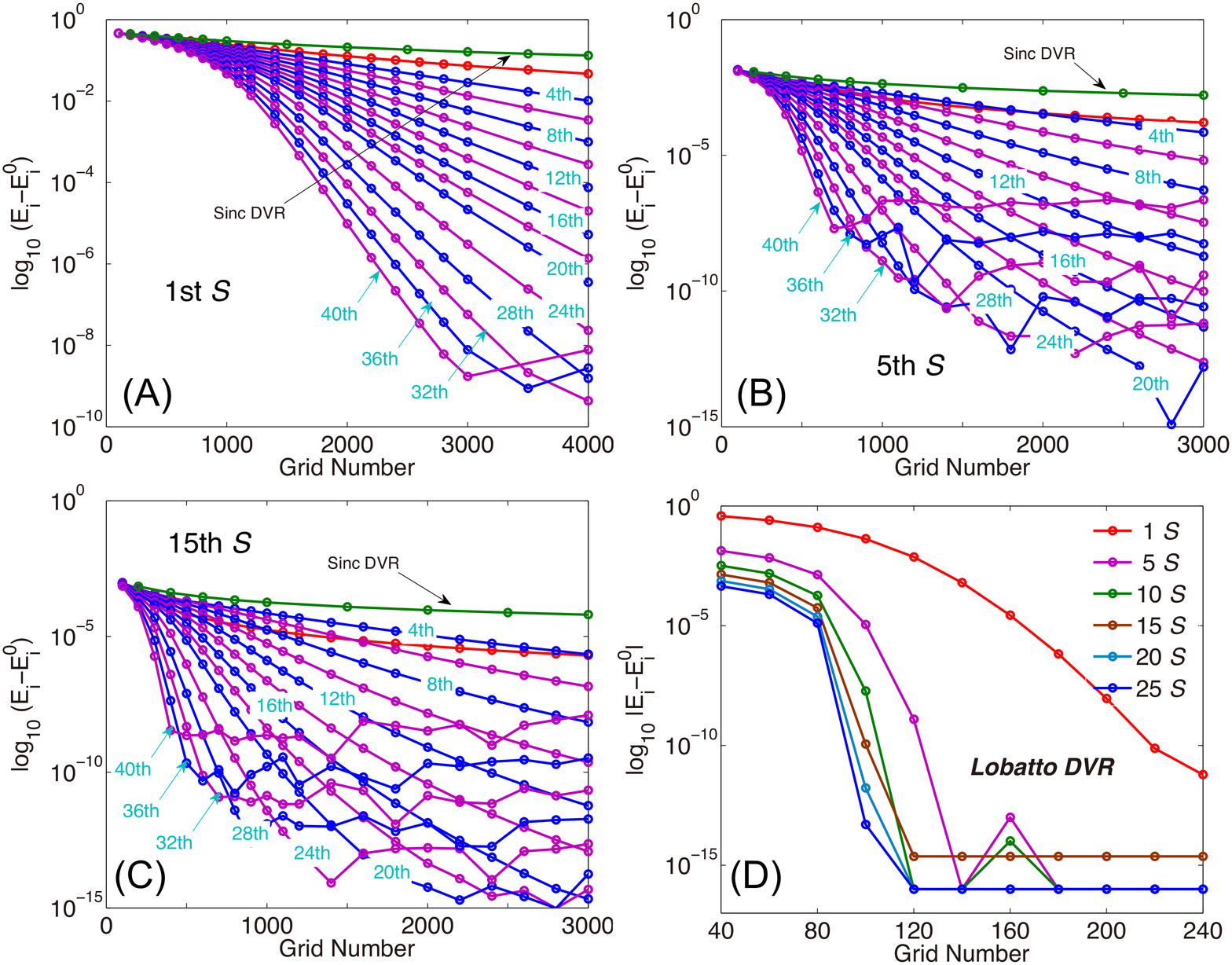}
\caption{\label{fig5} (A), (B) and (C): The convergence behaviors of energies of 1st, 5th and 15th electronic states of hydrogen atom with the higher order difference methods as a function of the number of grid points, along with the results using the Sinc-DVR method. (D): The convergence behaviors of energies of 1st, 5th, 10th, 15th, 20th and 25th electronic states of hydrogen atom as a function of numbers of grid points using the LDVR method in the same grid range as those used in the other three panels.} 
\end{figure}

\begin{figure}
\includegraphics[width=1.0\textwidth]{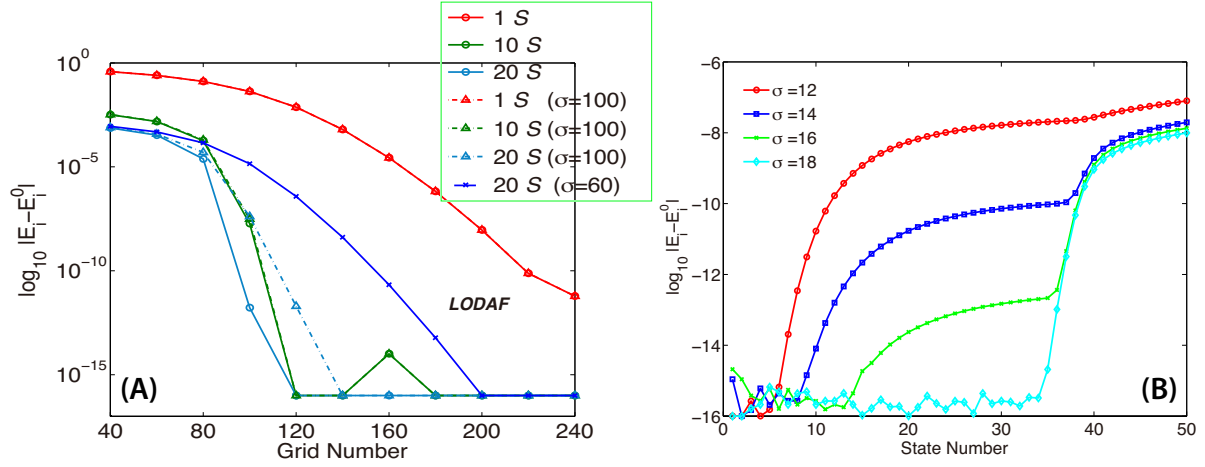}
\caption{\label{fig6} (A): The convergence behaviors of energies of the 1st, 10th and 20th electronic states of hydrogen atom as a function of total grid numbers, with fixed width $\sigma$=60.0 (only for 20th state, solid line with crosses) and 100a.u. of the weighting function for the DAF with the LDVR, along with the corresponding results with the LDVR (solid lines with circles). (B): The convergence behaviors of energies of low lying electronic states of hydrogen atom with different $\sigma$ for the DAF with the CDVR kernel.} 
\end{figure}

\end{document}